\documentclass[12pt,a4paper,final]{iopart}

\usepackage{iopams}  
\usepackage{graphicx}
\usepackage[breaklinks=true,colorlinks=true,linkcolor=blue,urlcolor=blue,citecolor=blue]{hyperref}

\begin{document}

\title{Nuclear Magnetometry Study of Spin Dynamics in Bilayer Quantum Hall Systems}

\author{M. H. Fauzi}
\address{Department of Physics, Tohoku University, Sendai
980-8578}
\ead{fauzi@m.tohoku.ac.jp}

\author{S. Watanabe}
\address{Institute of Science and Engineering, Kanazawa University, Kanazawa, 920-1192}
\ead{wshinji@se.kanazawa-u.ac.jp}

\author[cor1]{Y. Hirayama$^{1,2,3}$}
\address{$^1$Department of Physics, Tohoku University, Sendai
980-8578}
\address{$^2$ERATO Nuclear Spin Electronics Project, Sendai 980-8578}
\address{$^3$WPI-Advanced Institute for Materials Research, Tohoku University, Sendai 980-8577}
\eads{\mailto{hirayama@m.tohoku.ac.jp}}

\begin{abstract}

We performed a nuclear magnetometry study on quantum Hall ferromagnet with a bilayer total filling factor of $\nu_{\rm{tot}} = 2$. We found not only a rapid nuclear relaxation but also a sudden change in the nuclear spin polarization distribution after a one-second interaction with a canted antiferromagnetic phase. We discuss the possibility of observing cooperative phenomena coming from nuclear spin ensemble triggered by hyperfine interaction in quantum Hall system.

\end{abstract}

\vspace{2pc}

\section{Introduction}
The electron spins in the host GaAs semiconductor are coupled with the ensemble of nuclear spins mainly through the hyperfine (HF) interaction. The HF interaction lies at the heart of many fascinating phenomena including dynamic nuclear polarization, Knight shift, and Overhauser/Hyperfine field\cite{Abragam}, and is of importance to development of quantum computing based on hybrid ensemble of electron-nuclear spins. The interaction has been successfully utilized to probe and characterize various electron spin cooperative phenomena in 2D systems subjected to a strong magnetic field $B$ at low temperature, where the strong Coulomb interaction dominates the physics. Examples include evidence for the formation of a topological spin texture near a Landau level filling factor of $\nu=1$ as predicted by Shondhi {\it et.al} \cite{Sondhi93} via the Knight shift\cite{Barrett}, and nuclear spin relaxation $T_1$ time measurements\cite{Hashimoto2002}. Recently an canted antiferromagnetic state in a bilayer total filling factor $\nu_{\rm{tot}} = 2$ that supports linearly dispersing Goldstone modes\cite{Das97, Sarma, Hama} was experimentally verified by Kumada {\it et.al} who used the Knight shift\cite{Kumada07} and $T_1$ measurements\cite{Kumada06}.

A large portion of research has been devoted to the use of nuclear spins as a mere tool for studying electronic structures in quantum Hall systems, whereas little attention has been given as to how the electronic structures affect the nuclear spin. In fact, cooperative phenomena coming from an ensemble of nuclear spins induced by the HF interaction could lead to various interesting features including nuclear spin helix in 1D system\cite{Bernd} very recently observed in GaAs quantum wires\cite{Scheller}, or nuclear superradiance like effect in quantum dots\cite{Eto, Schuetz}. Yet to the best of our knowledge, the HF induced superradiance effect has not been observed to date. One of the strongest cooperative phenomena involving electron spins appears in a bilayer canted antiferromagnetic (CAF) state as evidenced by its very short nuclear spin relaxation $T_1$ time\cite{Kumada06}. Therefore, we expected there would be cooperative phenomena that could produce superradiance like effect when an ensemble of nuclear spins interacts selectively with the CAF state.

In this study, we developed a nuclear magnetometry and used it to demonstrate the possibility of collective nuclear spins relaxation due to interaction with the Goldstone mode of the CAF phase of total filling factor of $\nu_{\rm{tot}} = 2$ in the quantum Hall effect. We found that the initial number of polarized nuclear spins would affect the relaxation behaviour. A pump-probe technique performed at the spin transition at the filling factor of $\nu = 2/3$ was employed to dynamically polarize the nuclear spins and probe their relaxation dynamics\cite{Hashimoto2002}. We analyze the position of the spin transition to estimate the hyperfine field value and its full width at half maximum (FWHM) to qualitatively discuss the homogeneity of nuclear spin polarization in the well.

\begin{figure*}
\centering\includegraphics[width = 5in ]{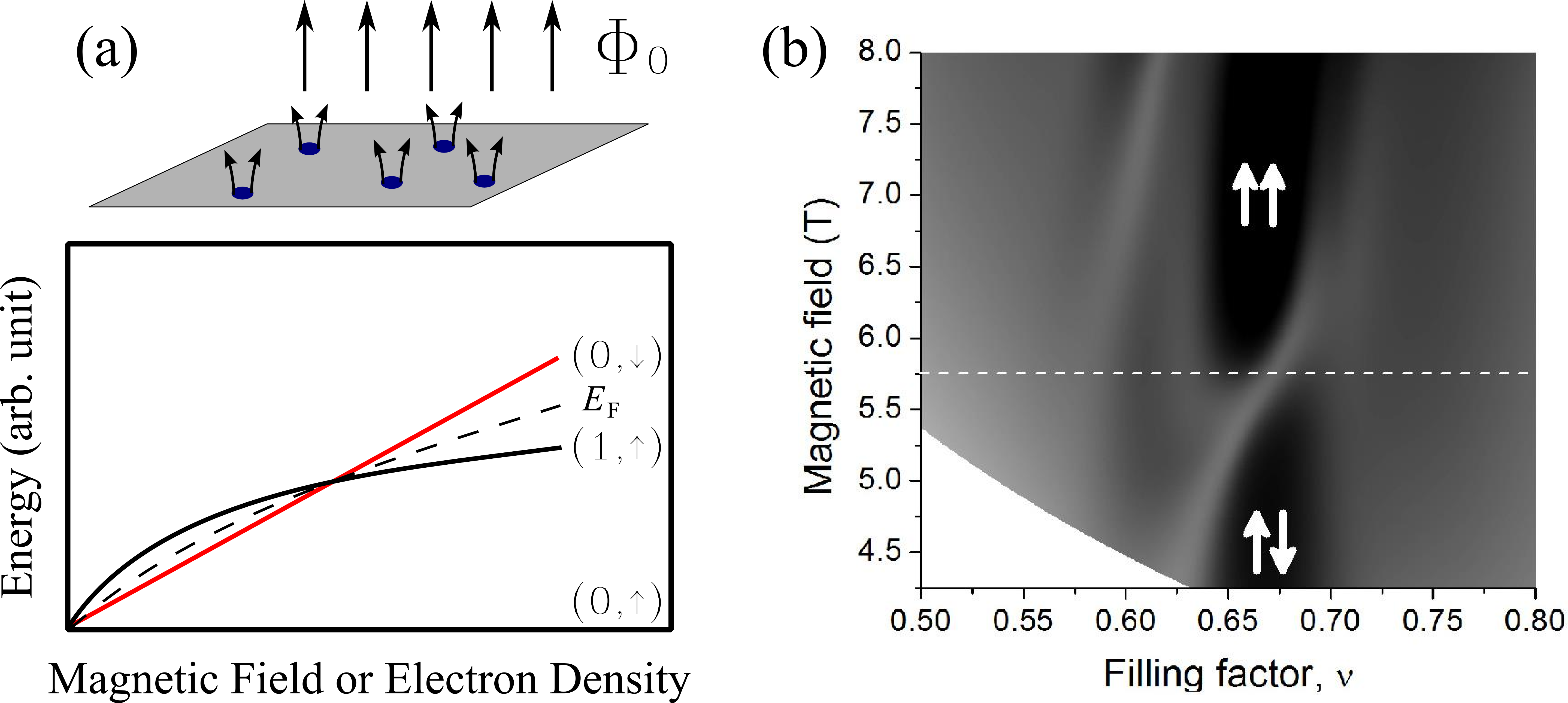}
\caption{\label{fig:wide} { (a) Composite fermion Landau level energy diagram at the filling fraction $\nu = 2/3$. The state is indicated by the bracket ($n, m$), $n$ corresponds to the number of Landau level $n = 0, 1, 2, ...$} and $m = \uparrow, \downarrow$ corresponds to the spin state $\pm 1/2$. (b) Two dimensional map of $R_{\rm{xx}}$ around filling fraction $\nu = 2/3$ obtained by scanning gate voltage and magnetic field. The spin configurations of each ground states are indicated by the white arrows. Dark (Bright) color has a low(high) resistance. Nuclear magnetometry discussed in section III and IV obtained at a fixed magnetic field of 5.75 T indicated by a horizontal white dashed line.}
\end{figure*}

\section{Spin Transition at $\nu = 2/3$}

In the composite fermion (CF) model, the filling factor $\nu = 2/3$ corresponds to integer filling factor $\nu_{\rm{CF}} = 2$ of CF, the two CF Landau levels are fully occupied below the Fermi energy $E_{\rm{F}}$. The CF particle comprises of one electron with two attached magnetic flux quanta\cite{Jain}. Consequently, the CFs experience a reduced magnetic field, which in mean field approximation, is given by 

\begin{equation}
B_{\rm{eff}} = B(1 - 2\nu)
\end{equation}
therefore an effective magnetic field experienced by the CFs at $\nu = 2/3$ is $B_{\rm{eff}} = -B/3$. Similar to the normal Landau level, the CFs energy spectrum is also quantized into a series CF Landau levels. Each CF Landau levels is separated by 

\begin{equation}
\hbar \omega_{\rm{cf}} = \frac{e \hbar}{m_{\rm{CF}}}\frac{B}{3}
\end{equation}

here $m_{\rm{cf}} \equiv \alpha \sqrt{B} m_0$ is the composite fermion effective mass \cite{Kukushkin}. Due to the Zeeman effect, each CF Landau levels further splits into two spin sublevels separated by $E_Z = |g^*|\mu_{\rm{B}}B$, where $\mu_{\rm{B}} = e\hbar/2m_0$ and $|g^*| = 0.44$. The levels $(0, \downarrow)$ and $(1, \uparrow)$ can cross each other at certain range of magnetic field $B$ when the Zeeman energy equals to the CF cyclotron energy gap (see Fig. 1a). The electronic system at the transition experiences a first-order spin transition from a spin-unpolarized ($\uparrow \downarrow$) to a spin-polarized ($\uparrow \uparrow$) ground state (see Fig. 1b) as the magnetic field increases above a critical field $B_{\rm{t}}$. Coexistence of two electron spin domains ($\uparrow \downarrow$ and $\uparrow \uparrow$) at the transition point has been uncovered through magnetotransport experiments\cite{Smet2001}, NMR spectroscopy\cite{Stern2004}, as well as from microscopic standpoint\cite{Verdene}. The field crossing at the spin transition is given by

\begin{equation}
\hbar \omega_{\rm{cf}} = \Delta_Z
\end{equation}

\begin{equation}
B_{\rm{t}} = \left(\frac{2}{3\alpha |g^*|}\right)^2
\end{equation}

For example assuming $\alpha = 0.633$, which is comparable to the theoretical value reported in ref\cite{Park}, yield $B_{\rm{t}} = 5.75$ T.

Now let's discuss the influence of nuclear spin polarization to the spin transition. GaAs has three isotopes namely $^{71}$Ga, $^{69}$Ga, and $^{75}$As with angular momentum $I = 3/2$. The interaction between an electron and nuclear spins in GaAs is mainly through the contact hyperfine interaction described by the following Hamiltonian

\begin{equation}
H = \frac{1}{2}A\left(I^+S^- + I^-S^+\right) + AI_ZS_Z
\end{equation}

Here $A$ is the hyperfine coupling constant. The first term describes the dynamical process between an electron spin and nuclear spin including dynamic nuclear polarization and nuclear polarization decay. The second term describes the influence of static hyperfine field $B_{\rm{N}} = A\langle I_{Z}\rangle/g^* \mu_{\rm{B}}$ produced by an ensemble of nuclear polarization on the electronic Zeeman energy.


\begin{equation}
\Delta_{\rm{Z}} = |g^*| \mu_{\rm{B}}(B + B_{\rm{N}})
\end{equation}

Accordingly, the field crossing at the spin transition in the presence of the hyperfine field $B_{\rm{N}}$ is

\begin{equation}
B_{\rm{t}} = \left (\frac{1 + \sqrt{1 - 9\alpha^2|g^*|^2B_{\rm{N}}}}{3\alpha |g^*|}\right)^2
\end{equation}

\begin{figure*}
\centering\includegraphics[width = 6.0in ]{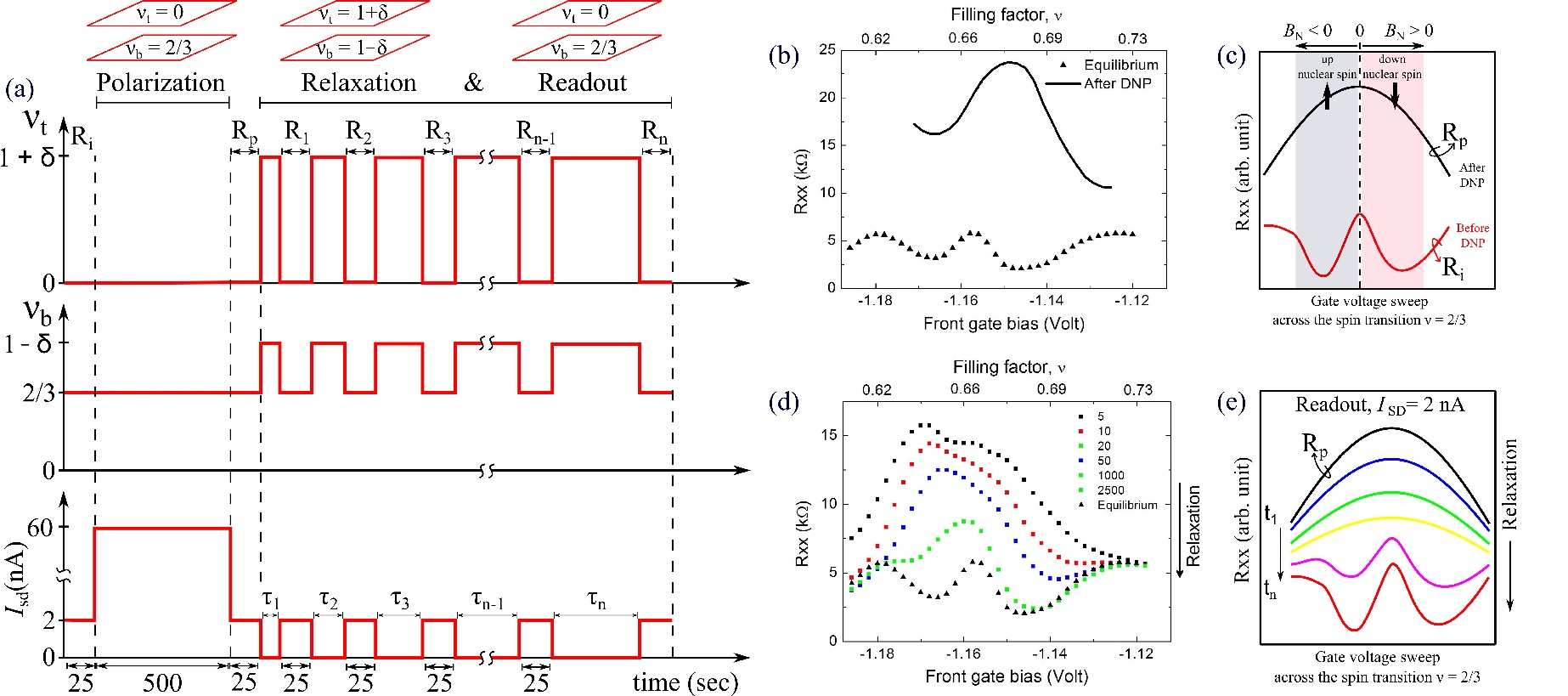}
\caption{\label{fig:wide} {(a) Schematic illustration of timing sequence diagram for the nuclear magnetometry measurement. $R_i$ and $R_p$ are acquisition sequences of spin transition profile before and after dynamic polarization respectively. $R_n$ ($n = 1, 2, 3, ...$) is acquisition sequence after interaction with electrons of the bilayer with exposure durations $\tau_n$. Acquisition time for each profile was $25$ seconds. (b) Spin transition profiles before (equilibrium) and after dynamic nuclear polarization and (c) its schematic cartoon. (d) Example of nuclear magnetometry measurement due to interaction with electrons of the bilayer: $\nu_{\rm{tot}} = 2$; $\delta = 0.37$; exposure durations $\tau$ of $5$ to $2500$ seconds and (e) a cartoon of the expected response for simple nuclear relaxation with equal depolarization rate.} }
\end{figure*}

From equation (7), one can estimate the hyperfine field $B_{\rm{N}}$ by monitoring the field crossing $B_{\rm{t}}$ as demonstrated in ref\cite{Li}. Alternatively, one can also estimate $B_{\rm{N}}$ from the spin transition peak position with fixed magnetic field by sweeping the gate voltage (electron density) across the transition. In fact, a tiny amount of the nuclear polarization down to $2 \%$ thermal equilibrium nuclear polarization can be detected sensitively by this spin transition\cite{Fauzi}.

\section{Nuclear Magnetometry Experimental Procedure}

Experiments were carried out on a high-quality 20-nm-wide bilayer GaAs quantum well separated by a 2.2 nm Al$_{0.3}$Ga$_{0.8}$As barrier. The energy separation of the symmetric and antisymmetric states, $\Delta_{\rm{SAS}}$, was $15$ K at the charge balanced condition for a total electron density of $n_{\rm{tot}} = 1.45 \times 10^{15}$ m$^{-2}$. The sample was patterned into a $30$-$\mu$m-wide Hall bar, and ohmic contact pads were made with Ni/AuGe/Ni alloys annealed at $420^0$ C. The carrier density of the top and bottom layers ($n_{\rm{f}}$ and $n_{\rm{b}}$) could be controlled independently from depletion to $4.0 \times 10^{15}$ m$^{-2}$ by applying bias voltages to the top gate made of a Ti/Au electrode deposited on top of the Hall bar and the $n^+$-GaAs substrate acting as the bottom gate. At a constant magnetic field $B$, the filling factor was tuned by controlling the carrier density in each layer. The total filling factor $\nu_{\rm{tot}} = \nu_{\rm{f}} + \nu_{\rm{b}}$ is the sum of the individual filling factors. All magnetotransport measurements were carried out using a lock-in technique at $13.4$ Hz and the sample was immersed in a mixture of He-3/He-4.

\begin{figure*}
\centering
\includegraphics[width = 6in ]{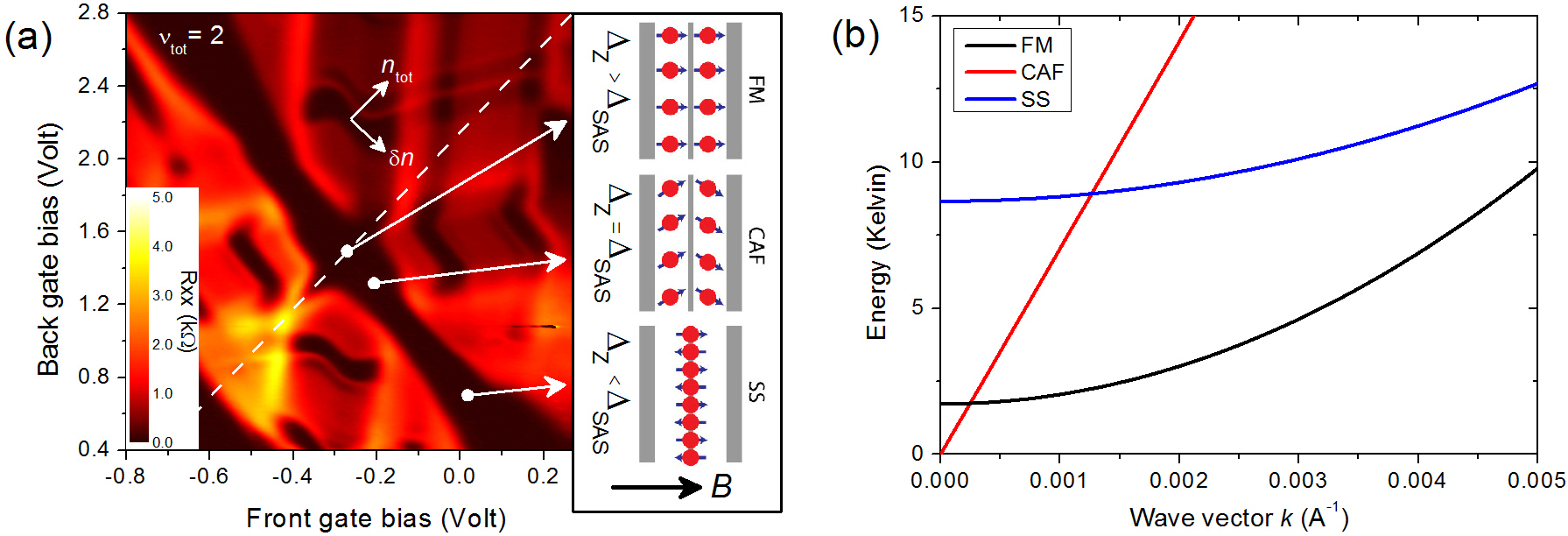}
\caption{\label{fig:wide} (a) Two dimensional plot of $R_{\rm{xx}}$ as a function of a back- and front-gate bias voltage highlighted along $\nu_{\rm{tot}} = 2$ with its possible spin configurations (see the box) for 5.75 T and 50 mK. The white dashed line corresponds to the zero charge imbalance $\delta = 0$. (b) Calculated dispersion energy curve for three different $\nu_{\rm{tot}} = 2$ bilayer states near $k = 0$. Parameters used in the calculation are $B = 5.75$ T, total density $n = 2.8 \times 10^{15}$ m$^{-2}$, intra-layer stiffness $J_s = 2.98$ K, inter-layer stiffness $J_s^d = 2.25$ K, inter-layer Coulomb energy $\epsilon_D^- = 63.82$ K, and exchange energy $\epsilon_X^- = 25.51$ K.}
\end{figure*}

The key aspect of our experimental protocol is schematically displayed in Fig. 2(a). First, the carrier density in one layer (bottom layer) was set to the spin transition at the filling factor of $\nu = 2/3$ at constant top $V_{\rm{tg}} = -1.157$ V and bottom gate bias voltages $V_{\rm{bg}} = +2.8$ V, and a constant magnetic field of $B = 5.75$ T. A high excitation current $I_{\rm{sd}} = 60$ nA was applied for $500$ sec (unless mentioned otherwise) to induce dynamic nuclear polarization (DNP). 

At the spin transition, the energy mismatches between electron and nuclear spin is reduced which allows them to couple effectively. The nuclear spin can be dynamically polarized effectively by the $I_{\rm{sd}} = 60$ nA current flow because when an electron spin scatters across two different domains, it does so by flipping a nuclear spin around the domain boundary to preserve total angular momentum. The spin transition profile before and after dynamic polarization were readout as $R_i$ and $R_p$, respectively, as displayed in Fig. 2(b). Spatial inhomogeneous nuclear spin polarization enhanced and broadened the spin transition peak. The enhancement to the left of the peak ($B_{\rm{N}} < 0$) is attributed to the upward nuclear spin polarization $\langle + I_{Z}\rangle$ and the right one ($B_{\rm{N}} > 0$) is attributed to the $\langle - I_{Z}\rangle$ as schematically shown in Fig. 2(c). Next, the carrier density was tuned in both layers to reach the quantum Hall state with $\nu_{\rm{tot}} = (1 + \delta) + (1 - \delta)$ for several sets of charge imbalance variables, $\delta$. The excitation current $I_{\rm{sd}}$ was turned off at this sequence. The polarized nuclear spins then interact with electrons of the bilayer with $\nu_{\rm{tot}} = 2$. We interrupted the process by temporarily restoring the filling factor to a one layer (the bottom layer) $\nu_{\rm{b}} = 2/3$ after a given interval of time "exposure time $\tau$" and the remaining nuclear polarization was readout as $R_n$ (n = 1, 2, 3, ...) by sweeping the filling factor across $\nu_{\rm{b}} = 2/3$ ($0.61 \to 0.73$) by varying the gate bias voltages at a measurement current $I_{\rm{sd}}$ of $2$ nA. Note that with this measurement current level, DNP is negligible. The top gate voltage sweep rate was $dV_{\rm{tg}}/dt \sim 3.52 \times 10^{-3}$ s$^{-1}$ at constant bottom gate voltage $V_{\rm{bg}} = +2.8$. The sweep time from $\nu_{\rm{b}} = 0.61$($V_{\rm{tg}} = -1.186,V_{\rm{bg}} = +2.8$) to $\nu_{\rm{b}} = 0.73$($V_{\rm{tg}} = -1.12,V_{\rm{bg}} = +2.8$) was about $25$ seconds and much faster than the nuclear spin relaxation time at $\nu = 2/3$ ($> 300$ seconds). The remaining nuclear polarization is reflected in the readout spin transition profile $R_n$. An example of complete measurement sequences is given in Fig. 2(d) for charge imbalance $\delta = 0.37$ where the resistance to the left and the right of the peak gradually decreased almost at the same rate towards its equilibrium shape. Fig. 2(e) shows a simplified situation in which both $\langle + I_{Z}\rangle$ and  $\langle - I_{Z}\rangle$ are depolarized equably. We will show you in the next section that this picture is no longer valid when an ensemble of nuclear spins interacts with the CAF state.

\begin{figure*}
\centering
\includegraphics[width = 6.0in]{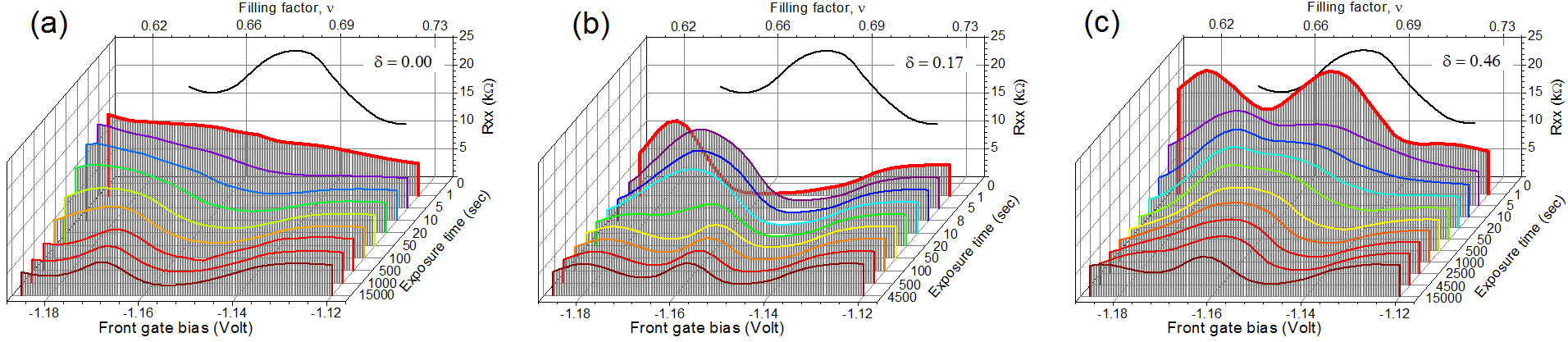}
\caption{\label{fig:wide} (a)--(c) Nuclear spin relaxation process, reflected in the spin transition evolution, due to interactions with (a) the ferromagnetic (FM) state $\delta = 0$, (b) the canted antiferromagnetic (CAF) state $\delta = 0.17$, (c) the spin single (SS) state $\delta = 0.46$. The solid black line in (a)--(c) is the initial nuclear spin polarization profile taken prior to interactions with electrons of the bilayer.}
\end{figure*}

\section{Results and discussions}

Figure 3(a) depicts a two-dimensional map of the longitudinal resistance $R_{\rm{xx}}$ highlighted along $\nu_{\rm{tot}} = 2$ as a function of the top- and the bottom-gate voltages at $5.75$ T and $50$ mK. The phase transition between different magnetic state along the $\nu_{\rm{tot}} = 2$ was driven by altering the normalized density imbalance $\delta \equiv (\nu_{\rm{t}} - \nu_{\rm{b}})/\nu_{\rm{tot}}$\cite{Brey}. The quantum Hall effect was preserved from the point of no charge imbalance $\delta = 0$, where the system possessed a ferromagnetic (FM) state, to a very large charge imbalance $\delta \approx 1$, where the spin configuration altered to a spin singlet (SS) state when the tunneling gap $\Delta_{\rm{SAS}}$ overwhelmed the Zeeman energy $\Delta_{\rm{Z}}$\cite{Kumada06}. Level crossing did not take place because the transition from FM to SS phases occured through two second-order phase transition via an intermediate state, namely the canted antiferromagnetic (CAF) state\cite{Das97, Sarma, Hama}.

According to ref\cite{Hama}, the lowest dispersion energy for ferromagnetic, canted, and spin singlet can be calculated analytically by the following equations

\begin{equation}
E_{\rm{FM}}(k) = \frac{4J_s}{n}k^2 + \Delta_Z
\end{equation}

\begin{equation}
E_{\rm{CAF}}(k) =  |k| \sqrt{\frac{8J_s^d}{n}\left(\frac{2J_s}{n}k^2 + 2\epsilon_D^- - 2\epsilon_X^- \right)}
\end{equation}

\begin{equation}
E_{\rm{SS}}(k) = \frac{2J_s^d}{n}k^2 + \frac{\Delta_{\rm{SAS}}}{2\sqrt{1 - \delta^2}}
\end{equation}

here $J_s$, $J_s^d$, $\epsilon_D^-$, and $\epsilon_X^-$ are intra-layer stiffness, inter-layer stiffness, inter-layer Coulomb energy and exchange energy, respectively, with the explicit formula is given in ref\cite{Hama}. The dispersion curve near $k = 0$ for those three different states is displayed in Fig. 3(b). It is clear that the CAF state is the only gapless mode and it has a linear dispersion curve near $k = 0$, while the other two states have non zero excitation gap at $k = 0$ with $E_{\rm{FM}} = 1.725$ K, and $E_{\rm{SS}} = 8.66$ K.

Now let us analyze the nuclear spin relaxation due to interactions with the electrons of the bilayer $\nu_{\rm{tot}} = 2$. In the SS state depicted in Fig. 4(c), the time required to reach the equilibrium spin transition shape, $T_{\rm{eq}}$, is very long ($T_{\rm{eq}} > 4500$ sec). In addition, the way that the spin transition curve relaxes is qualitatively quite similar to our expected relaxation behavior in Fig. 2(e). This is not surprising since the nuclear subsystem is well isolated from the electronic system, the nuclear Zeeman energy ($\sim$ MHz) is several order of magnitude that the excitation gap of the SS state. It is therefore reasonable that the exchanges of energy and angular momentum are very inefficient and the SS state cannot contribute much to the nuclear spin relaxation process. The relaxation channel is mainly governed by nuclear spin diffusion. For the FM state shown in Fig. 4(a), the measured $T_{\rm{eq}}$ was $\sim 500$ sec, which is almost ten times faster than in the SS state. The shape of the spin transition towards equilibrium revealed that it was rather distinct from the previous one measured in the SS state. The curve's fall was asymmetrically; that is the resistance to the right of the peak, corresponding to $B_{\rm{N}} > 0$, dropped more than it did on the left.

The most striking feature of the nuclear spin relaxation process was observed in the CAF state depicted in Fig. 4(b). First of all, the $T_{\rm{eq}}$ was very short, $\sim 50$ sec. This indicates the appearance of electron spin fluctuations which have a high spectral density at the Larmor nuclear frequencies of Ga and As\cite{Girvin}; this is suggestive of the linearly dispersion Goldstone mode as displayed in Fig. 3(b). What makes the relaxation process even more interesting is that the initial characteristics of the DNP completely disappeared one second after exposure to the CAF state. The broad transition curve suddenly became very narrow. The resistance to the right of the peak dropped to almost zero because the downward nuclear spin polarization, which is higher in energy than the upward one, completely relaxed. The spin transition curve moved rapidly back to equilibrium by shifting back to a higher filling factor at $\nu \approx 0.66$ within $\tau \approx 50$ seconds, while its width remained narrow during the evolution. The observed response clearly indicated a sudden change in the nuclear spin polarization distribution after one second of interaction with the CAF state. From the peak shifting and matching condition between the Zeeman and Coulomb energy scales\cite{Akiba}, we can estimate the hyperfine field $B_{\rm{N}}$ from the remaining nuclear spin polarization one second after exposure to the CAF state. The estimated $B_{\rm{N}} = \sqrt{B_{1}B_{2}} - B_{1}$ is approximately $0.66$ T. It roughly corresponds to a $12.5\%$ spatially homogeneous nuclear spin polarization, assuming that if all GaAs nuclear spins were fully polarized, the hyperfine field would be as high as $5.3$ T\cite{Paget}. Here, $B_{2} = 4.5$ T is where the filling factor of the spin transition approximately coincided with the transition at $B_{1} = 5.75$ T one second after exposure to the CAF state (see Fig. 1(b)).

\begin{figure*}
\centering\includegraphics[width = 5in ]{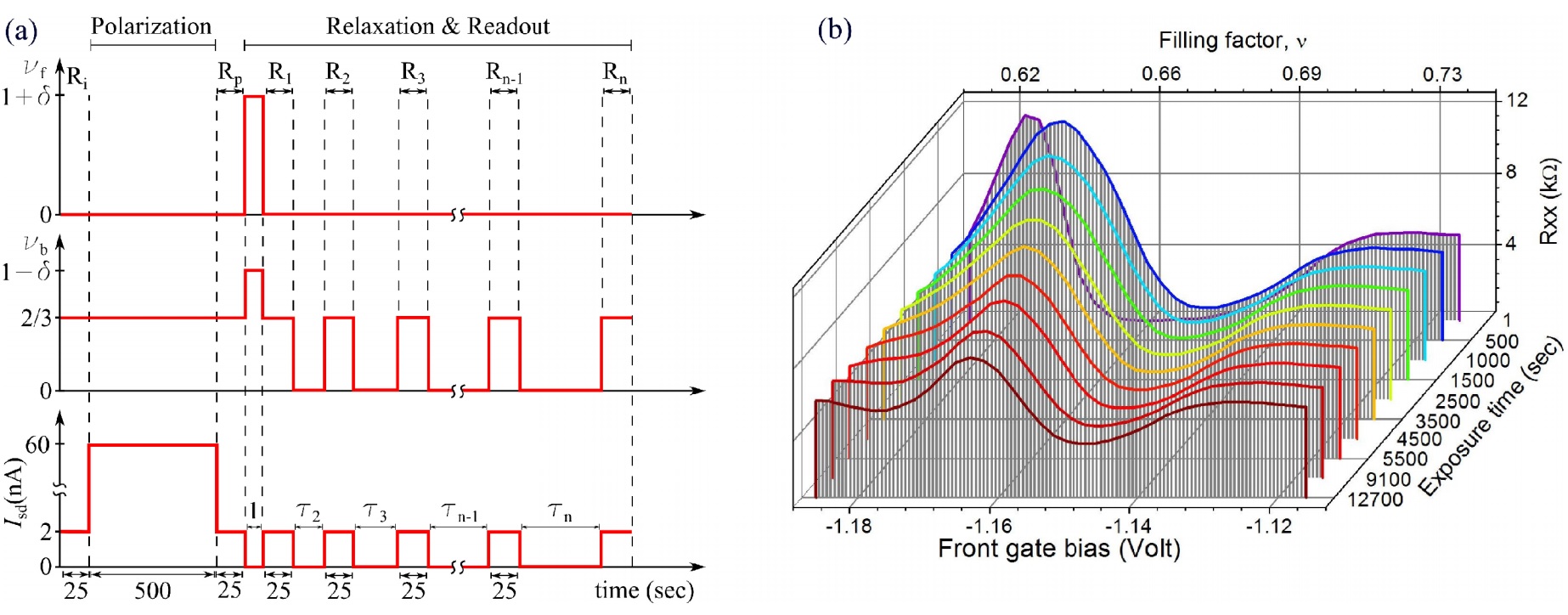}
\caption{\label{fig:wide} (a) Extended experimental protocol and timing sequence to study a sudden redistribution of nuclear spin polarization due to interaction with the CAF state. (b) The spin transition profile due to interaction with the CAF state for $\tau = 1$ second and no electrons for $\tau > 500$ seconds. }
\end{figure*}

To get further insight into a sudden redistribution of the nuclear spins ensemble due to interaction with the CAF state for $\tau = 1$ second, we carried out the experimental procedure described in Fig. 5(a). The procedure in principle was carried out similarly to the previous procedure in Fig. 2(a). The only difference was that after readout sequence $R_1$ finished, instead of setting the total filling factor back to $\nu_{\rm{tot}} = 2$ again, we depleted the electrons in each layer for a given interval of time before the next readout sequence ($R_2$, $R_3$, ..., $R_n$) was started. Each layer was completely depleted by applying $V_{\rm{tg}} = -1.1$ Volt to the top gate and $V_{\rm{bg}} = -0.5$ Volt to the bottom gate. Apparently, the evolution of the spin transition shown in Fig. 5(b) after depletion followed much the same way as in Fig. 4(b), however, with a longer time scale of about two order of magnitude slower to reach equilibrium. The result suggests that the redistribution is solely due to interaction with the CAF state. Another interesting point is that we could keep the remaining homogeneous nuclear polarization for a significant time period (about $500$ seconds) by just depleting the electrons in each layer one second after exposure to the CAF state.

Let us elaborate on how the response changes when we decrease the current pumping time from $P = 500$ seconds. The number of polarized nuclear spins would decrease as a result of shortening the current pumping time. For all of the data presented in Fig. 6, the exposure time to the CAF state with $\delta = 0.17$ was fixed to $\tau = 1$ second. Evidently, the response in terms of the spin transition's position and width showed a dependence on the current pumping time. The largest shift in the spin transition's position with respect to the equilibrium position ($\nu \approx 0.66$) appeared at $P = 500$ seconds, and it decayed with decreasing polarization time (see the black arrows in Fig. 6(a)). Interestingly as depicted in figure 6(b), its width had the opposite tendency; i.e., its value peaked at the shortest polarization time and became narrower with increasing polarization time. This suggests that for $P \le 200$ seconds, the downward nuclear spin polarization did not completely relax.

The observed response might be the fingerprint of an emission due to a collective nuclear spin relaxation, in analogy with superradiance emission in quantum optics\cite{Dicke}. The possibility of observing superradiance emissions from an ensemble of nuclear spins in a magnetic field was first put forward by Bloembergen and Pound\cite{Bloembergen} and experimentally observed by Kiselev {\it et al} \cite{Kiselev}.  The motions of the inverted polarized nuclear spins can become highly correlated when the resonance circuit frequency matches the Larmor frequency of nuclei. Indeed, the present system is fundamentally different than the one studied by Bloembergen and Kiselev {\it et al}. In our case the nuclear spins interact mainly with the electron spins via the HF coupling but not with the resonance circuit as in their case. Indeed nuclear superradiance could also induced via the HF coupling as pointed out in ref\cite{Eto, Schuetz}. This type of superradiance effect needs several prerequisite conditions below that can be satisfied in our case:

\renewcommand{\theenumi}{\roman{enumi}}%
\begin{enumerate}
  \item A strong HF coupling is needed and it has to be much stronger than the nuclear dipolar coupling. For GaAs case, the HF coupling strength is about $90$ $\mu$eV\cite{Paget} which is several order of magnitude larger than that of the nuclear dipolar coupling.
  \item It requires direct flip-flop process as described in the first term of equation (5). At high magnetic field, the angular momentum exchange is usually impeded due to large electronic Zeeman energy. However, this requirement is easily met for the CAF state since it has gapless excitation energy at $k = 0$.
  \item Electrons have to be strongly correlated. In fact, the coherence length of the Goldstone mode is very long and hence could satisfy the requirement.
  \item Lastly we need to stress that although the Goldstone mode has continuous energy dispersion, it has a strong spectral density at the Larmor frequency of Ga and As nuclear species as evidence by a very short $T_1$ time.
\end{enumerate}

\begin{figure*}
\centering
\includegraphics[width = 5in]{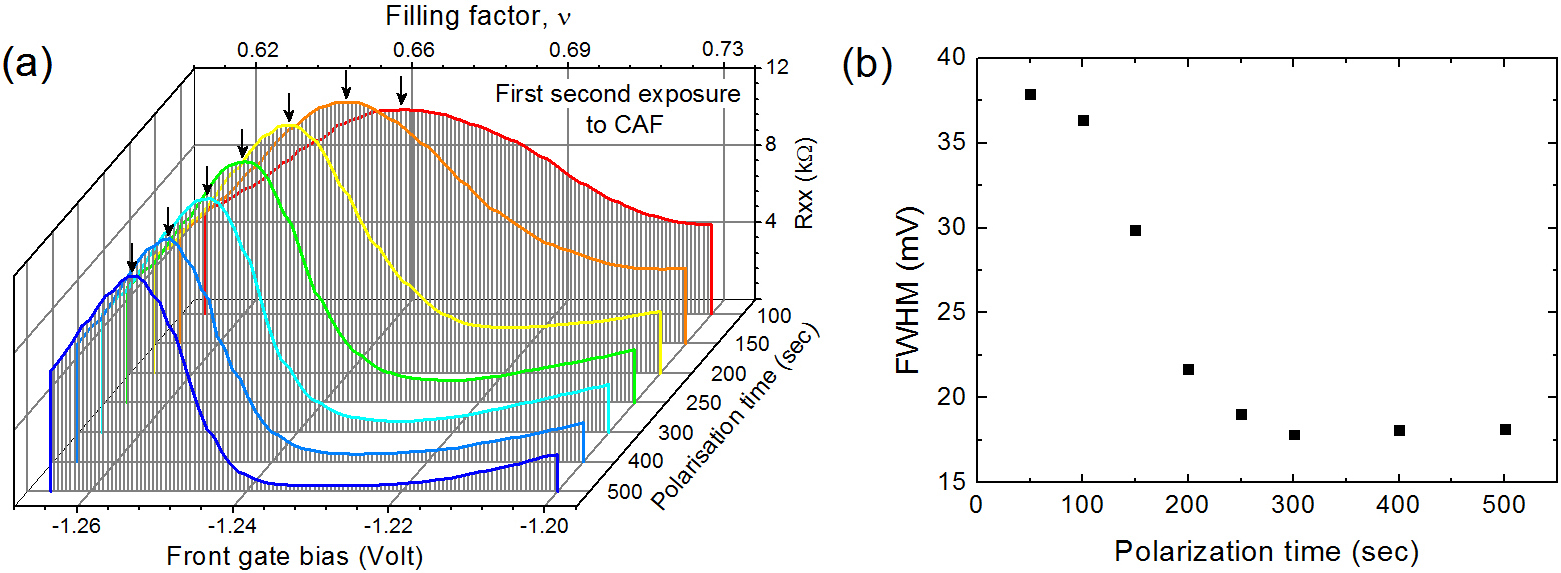}
\caption{\label{fig:wide} (a) The evolution of the spin transition after one second of exposure to the CAF phase $\delta = 0.17$ as a function of polarization time $P$ ranging from $100$ to $500$ seconds. The black arrow indicates the spin transition positions which depend on the hyperfine field $B_{\rm{N}}$ at a fixed applied magnetic field $B$. (b) The full width at half maximum (FWHM) as a function of polarization time extracted from Fig. 6(a).}
\end{figure*}

The detailed spin transition curve $\tau = 1$ second after the exposure could help us to identify the presence of special electronic states associated with broken planar symmetry. Figure 7 illustrates the nuclear relaxation process for a certain range of temperatures and/or charge imbalances and plots the extracted FWHM. The characteristic response associated with the CAF state for $\delta = 0.24$ disappeared upon increasing the sample temperature from $50$ to $200$ mK (see Fig. 7(a)--(b)). The spin transition curve noticeably became very broad for a $\tau = 1$ second exposure. This could be interpreted as possibly a straightforward signature of the transition between the CAF and SS phases. The long-ranged ordering was destroyed, resulting in incoherent coupling between the nuclear and electron spins. We note that the difference in the final transition position at equilibrium ($\tau = 1000$ seconds) between the $50$ and $200$ mK data sets is due to a decrease in the thermal equilibrium nuclear spin polarization of about $3.2 \%$ at $B = 5.75$ T.

Depicted in Fig. 7(c) is the FWHMs extracted from a set of $\delta$ values at $50$ mK. All data were extracted from $\tau = 1$ second responses except for those $\delta < 0.1$, which are from $\tau = 20$ seconds responses and can be used as lower bounds for the $\tau = 1$ second values. The transitions between the different electron spin phases are clearly marked by sudden changes in the width of the transition curve at $\delta = 0.104$ and $\delta = 0.276$ (indicated by the vertical red dashed lines). The clear transition helped us to construct the thermodynamic phase-diagram depicted in Fig. 7(d). Although the data were limited to the range of $50$ to $200$ mK, we can see that the area at which the CAF state was expected to occur shrank as the sample temperature was raised. We estimated that the CAF state would completely disappear above $300$ mK by extrapolating the data linearly to the point where both lines converge (blue oval). For a quantum Hall state with easy-plane quantum ferromagnets, this point is associated with Kosterlitz-Thouless (KT) transition temperature\cite{Kosterlitz}. The estimated $T_{\rm{KT}} \sim 300$ mK from our experiment is lower than the theoretical prediction $T_{\rm{KT}} = 1$ K\cite{Sarma}, but in agreement with the previously estimated $T_{\rm{KT}}$ deduced from resistively detected $T_1$ measurements\cite{Kumada06}. We believe that this discrepancy was due to disorder that might significantly lower the critical temperature $T_{\rm{KT}}$\cite{Sun}.

\section{Summary}
In summary, we uncovered an unusual nuclear spin relaxation process due to interaction with the CAF state by measuring the full profile of the $\nu = 2/3$ spin transition. We observed that only when the current pumping time was greater than the $200$ seconds did the downward nuclear spin polarization completely relax after a one second interaction with the CAF state. This could indicate the possibility of a collective relaxation from a large ensemble of polarized nuclear spins. Our nuclear magnetometry scheme and analysis of the FWHM of the spin transition $\nu = 2/3$ could be used to identify the transition between different phases existing at $\nu_{\rm{tot}} = 2$ and draw a $T-\delta$ diagram of the CAF state. 

\begin{figure*}
\centering
\includegraphics[width = 5in ]{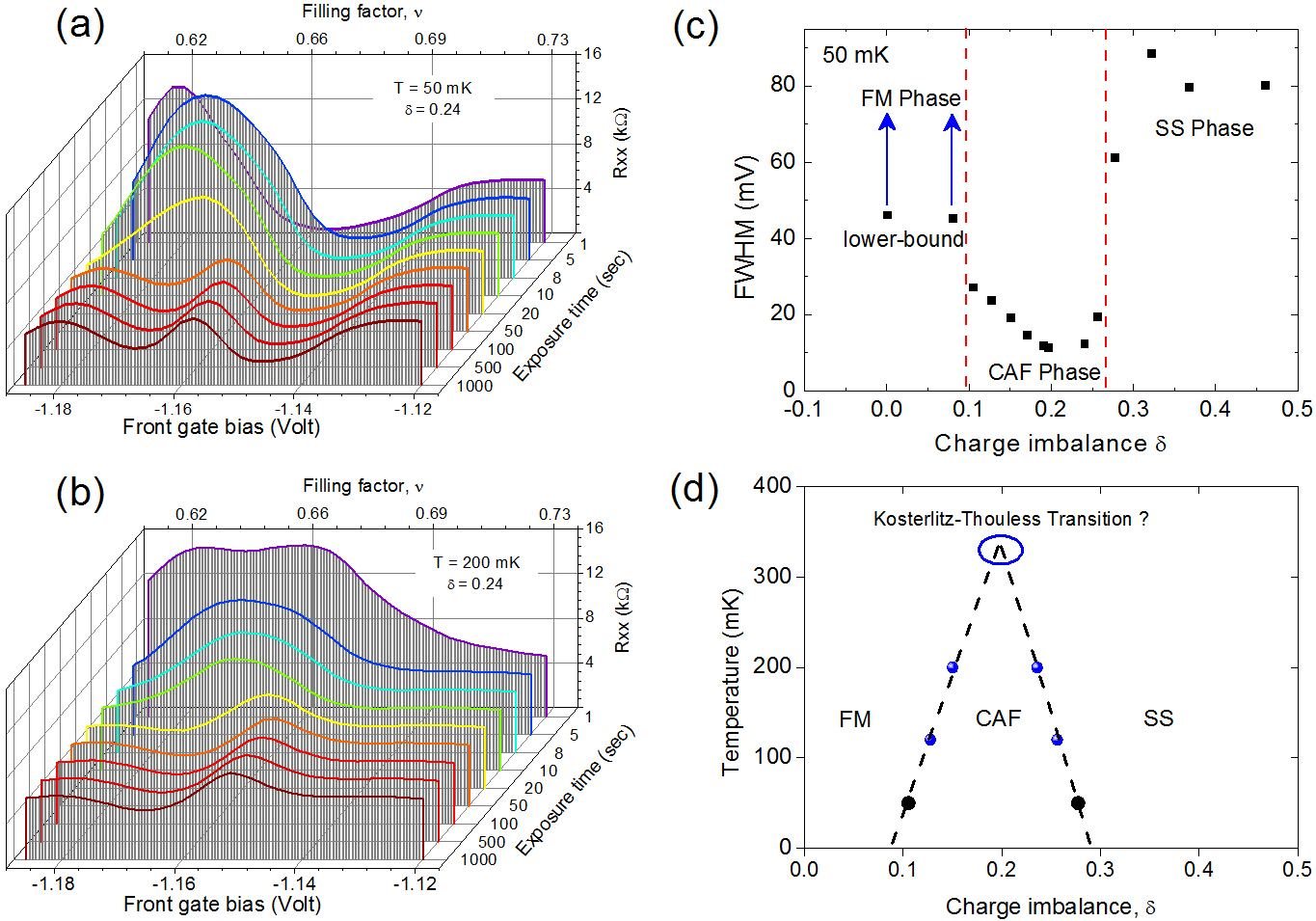}
\caption{\label{fig:wide} (a)--(b) Nuclear spin relaxation process exposed to the bilayer total filling factor $\nu_{\rm{tot}} = 2$ with $\delta = 0.24$ at $50$ and $200$ mK. (c) Plot of FWHM a one second after exposure to the bilayer total filling factor $\nu_{\rm{tot}} = 2$ with varying $\delta$ at $50$ mK. (d) Phase-diagram $\nu_{\rm{tot}} = 2$ as a function of charge imbalance $\delta$ and temperature. The 50 mK data points (black dots) are extracted from figure 6(c).}
\end{figure*}

{\bf Acknowledgements.}
We thank K. Muraki of NTT Basic Research Lab for providing us the bilayer wafer. We thank Y. Hama, K. Akiba, S. Miyamoto, K. Hashimoto, and T. Hatano for fruitful discussions. We gratefully acknowledged the partial financial support by the Tohoku University GCOE program.

\end{document}